\documentclass[conference,onecolumn]{IEEEtran}
\IEEEoverridecommandlockouts
\usepackage{cite}
\usepackage{amsmath,amssymb,amsfonts}
\usepackage{algorithmic}
\usepackage{graphicx}
\usepackage{url}
\usepackage[table]{xcolor}
\usepackage{booktabs}
\usepackage{multirow}
\usepackage{textcomp}
\usepackage{balance}
\usepackage{xcolor}
\usepackage{tcolorbox}
\usepackage{xspace}

\newcommand{\reqinone}{{\sc ReqInOne}\xspace}

\usepackage{tikz}
\newcommand{\circled}[1]{%
  \tikz[baseline=(char.base)]{
    \node[shape=circle,draw,inner sep=0.5pt,fill=black,text=white] (char) {#1};}}

\def\BibTeX{{\rm B\kern-.05em{\sc i\kern-.025em b}\kern-.08em
    T\kern-.1667em\lower.7ex\hbox{E}\kern-.125emX}}
\begin{document}

\title{\reqinone: A Large Language Model-Based Agent for Software Requirements Specification Generation\\}

\author{
  Taohong Zhu\IEEEauthorrefmark{1},
  Lucas C. Cordeiro\IEEEauthorrefmark{1},
  Youcheng Sun\IEEEauthorrefmark{2}\\
  \IEEEauthorrefmark{1}Department of Computer Science, The University of Manchester, Manchester, UK\\
  \{taohong.zhu, lucas.cordeiro\}@manchester.ac.uk \\
  \IEEEauthorrefmark{2}Mohamed bin Zayed University of Artificial Intelligence, Abu Dhabi, UAE\\
  youcheng.sun@mbzuai.ac.ae
}

\maketitle

\begin{abstract}
Software Requirements Specification (SRS) is one of the most important documents in software projects, but writing it manually is time-consuming and often leads to ambiguity. Existing automated methods rely heavily on manual analysis, while recent Large Language Model (LLM)-based approaches suffer from hallucinations and limited controllability. In this paper, we propose \reqinone, an LLM-based agent that follows the common steps taken by human requirements engineers when writing an SRS to convert natural language into a structured SRS. \reqinone adopts a modular architecture by decomposing SRS generation into three tasks: summary, requirement extraction, and requirement classification, each supported by tailored prompt templates to improve the quality and consistency of LLM outputs.

We evaluate \reqinone using GPT-4o, LLaMA 3, and DeepSeek-R1, and compare the generated SRSs against those produced by the holistic GPT-4-based method from prior work as well as by entry-level requirements engineers. Expert evaluations show that \reqinone produces more accurate and well-structured SRS documents. 
The performance advantage of \reqinone benefits from its modular design, and experimental results further demonstrate that its requirement classification component achieves comparable or even better results than the state-of-the-art requirement classification model.

\end{abstract}


\section{Introduction}
Requirements engineering is a critical phase in software development, ensuring stakeholder needs are accurately captured and translated into implementable specifications. The Software Requirements Specification (SRS) is the primary outcome of requirements engineering, which defines the expected functionality, constraints, and operational environment of a software system \cite{hofmann2001requirements}. A high-quality SRS must be unambiguous, complete, consistent, and traceable to guide design, implementation, and testing \cite{doe2011recommended}. However, producing such specifications is challenging due to the reliance on natural language, which often leads to vagueness, contradictions, and increased communication overhead between stakeholders \cite{belfo2012people}. Tools like Visual Paradigm \cite{VisualParadigm}, ReqView \cite{reqview}, and Elementool \cite{elemtool} offer templates and diagrams but still require extensive manual input. NLSSRE \cite{georgiades2010automatic} automates requirement extraction but does not support generating complete SRS content such as use cases and glossaries.

Recent advances in Large Language Models (LLMs) offer opportunities to automate and enhance requirements engineering tasks. Trained on vast real-world data, LLMs can generate human-like text with billions of parameters \cite{chen2021evaluating, kasneci2023chatgpt, zhao2023survey}. With the emergence of models such as LLaMA \cite{roziere2023code}, GPT \cite{achiam2023gpt}, and DeepSeek \cite{guo2025deepseek}, techniques like zero-shot \cite{radford2019language}, few-shot \cite{brown2020language}, and chain-of-thought \cite{wei2022chain} prompting have significantly improved model performance across diverse tasks. In the context of requirements engineering, LLMs have been applied to classify requirements \cite{hey2020norbert, surana2019intelligent}, evaluate user story quality \cite{ronanki2022chatgpt}, assess completeness \cite{luitel2024improving}, and support information extraction and architectural design \cite{ronanki2023investigating, ezzini2023ai, abdelfattah2023roadmap, el2023ai}.
In \cite{endres2024can}, LLMs are employed to translate informal code comments into formal postcondition assertions.
The framework in \cite{leite2024extracting} leverages ChatGPT and counterexample-guided refinement to automatically extract and verify formal postconditions for Ethereum smart contract functions based on natural language descriptions. \cite{krishna2024using} designed a prompt to enable LLMs to directly generate a full SRS from natural language text. However, an SRS includes multiple sections, the task involves not only converting natural language into well-structured requirements but also correctly placing each one into the appropriate section. This makes SRS generation task more complex. Directly prompting an LLM to generate the full SRS can lead to hallucinations, underscoring the need for a suitable transformation strategy and well-crafted prompts \cite{chang2024survey, fan2023large}.


This paper introduces \reqinone, an LLM-based agent designed to automatically convert natural language texts, such as stakeholder requirements, meeting transcripts, and conversational records, into a structured SRS. \reqinone consists of three core components: Summary Task, Requirement Extraction Task, and Requirement Classification Task, each guided by a tailored prompt template to perform its specific role in the SRS generation process. By coordinating these components, \reqinone efficiently produces well-structured SRS documents from unstructured text.

Furthermore, we construct ReqFromSRS, a dataset consisting of 100 functional requirements and 100 non-functional requirements, manually extracted from 22 real-world SRS documents in the PURE dataset \cite{ferrari2017pure}. We release all prompt templates, code, datasets, generated SRSs, and experimental results in Github \cite{ReqInOne} to support future research.

\begin{figure*}[h]
    \centering
    \includegraphics[width=\textwidth]{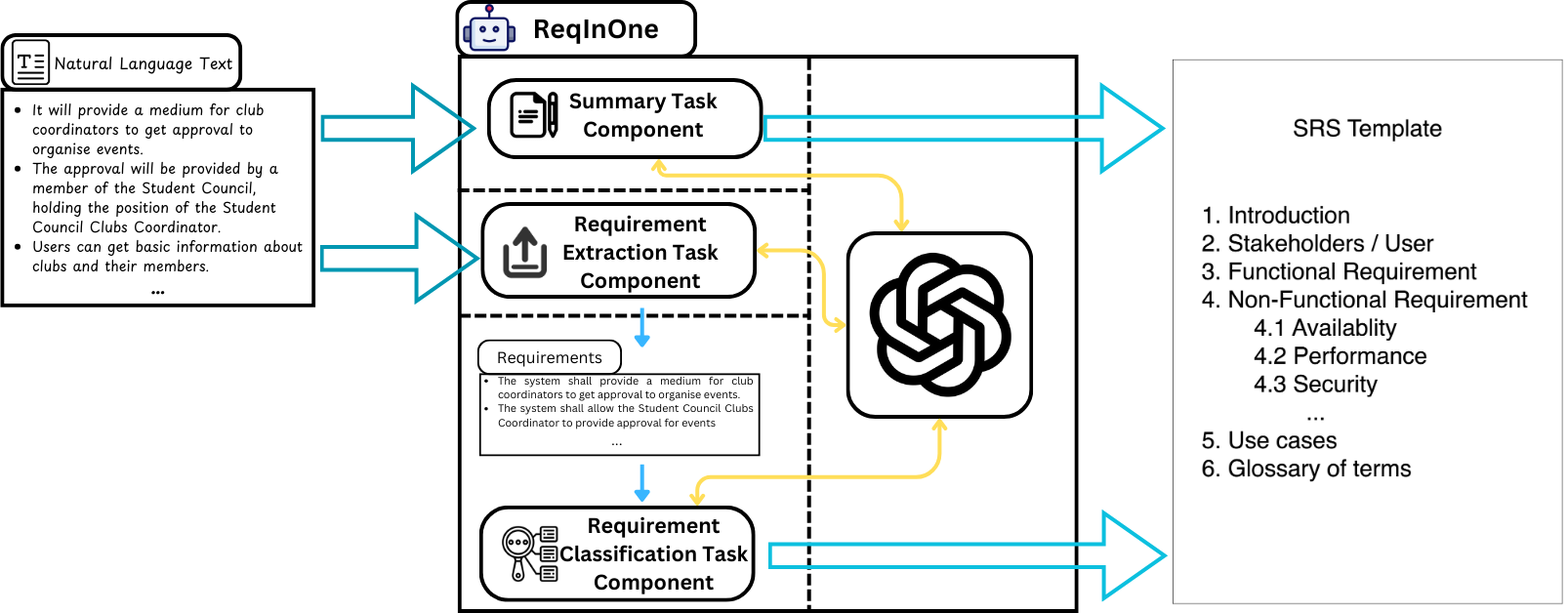}
    \caption{Overview of the \reqinone}
    \label{fig:overview}
\end{figure*}

\section{METHODOLOGY}

\subsection{Overview of \reqinone}
Prior studies have shown that LLMs may struggle to effectively handle complex tasks when prompted to complete them in a single step. However, when guided by a chain-of-thought approach \cite{wei2022chain}, such tasks can often be decomposed into a series of simpler sub-tasks, enabling more reliable and accurate performance. For instance, Tian et al. demonstrated that LLMs are less effective at directly repairing buggy code compared to scenarios where the error location is first identified by either human annotators or external tools, and the LLM is then responsible solely for repairing the localized snippet \cite{tian2024evaluating}. Inspired by this finding, we hypothesize that the task of converting natural language into a structured SRS, rather than being executed as a single-step transformation as in the work of Krishna et al. \cite{krishna2024using}, can similarly benefit from decomposition into a set of simpler, more focused sub-tasks. By adopting this multi-step strategy, we aim to improve LLM performance and generate higher-quality SRS outputs.

Building on this hypothesis, we design \reqinone to follow the common steps taken by human requirements engineers when converting natural language text into a structured SRS. Rather than treating the conversion as a single-step process, \reqinone decomposes the task into three sub-tasks: Summary, Requirement Extraction, and Requirement Classification. To efficiently handle these sub-tasks, \reqinone consists of three specialized components, each responsible for executing a specific task. Through task scheduling and coordination among these components, \reqinone processes natural language input and generates a well-structured SRS.

The overall workflow of \reqinone is illustrated in Figure \ref{fig:overview}. It begins with an input of verbose natural language text, typically reflecting stakeholder needs and high-level system requirements. \reqinone first invokes the Summary Task component, which processes the input text for summarization. The Summary Task component prompts the LLM based on the natural language text and utilizes the output of the LLM to populate the SRS template. The generated content is used to populate specific sections of the SRS template. As an example of the SRS template shown in Figure \ref{fig:overview}, the Summary Task component is responsible for filling in summary-type sections of the SRS template, including the Introduction, Stakeholders/Users, Use Cases, and Glossary sections. Additionally, the SRS template illustrated in Figure \ref{fig:overview} is merely an example, adapted from the template used in \cite{krishna2024using}, which in turn is based on IEEE specifications \cite{ieee830-1998}. Users are free to adopt any other SRS template that suits their specific needs. To do so, they simply need to identify the summary-type sections within their chosen template and accordingly modify the prompt template used in the Summary Task component (details in \ref{sec: Summary Task Component}).

Following this, \reqinone proceeds to the Requirement Extraction Task component, which extracts structured requirements from the natural language text. The Requirement Extraction Task component prompts the LLM based on the natural language text, and the output of the LLM constructs a list of structured requirements extracted from natural language text. This list of extracted requirements is then passed as input to the Requirement Classification Task component for further processing.

Once the list of extracted requirements is generated, it is passed to the Requirement Classification Task component for categorization. This component is responsible for classifying the requirements into Functional Requirements (FRs) and Non-Functional Requirements (NFRs). Moreover, the Requirement Classification Task component further refines the classification of NFRs by assigning them to specific subtypes such as availability, performance, security, and other relevant NFR subtypes. The Requirement Classification Task component prompts the LLM to classify requirements and determines their category based on the output of the LLM.

The classified requirements are then populated into the corresponding sections of the SRS template. FRs are placed in the FRs section, while NFRs are categorized into subsections under NFRs section. Once all three components complete their respective tasks, the conversion of natural language input into a structured SRS is finalized. Through this coordinated execution of its three core components, \reqinone provides an efficient and automated solution for generating SRS documents.

\subsection{Summary Task Component}
\label{sec: Summary Task Component}

\begin{figure}[htbp]
    \centering
    \includegraphics[width=0.6\linewidth]{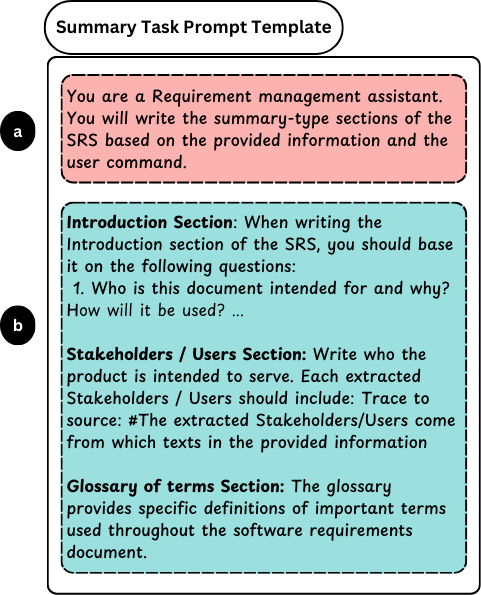}
    \caption{Prompt Template for Summary Task}
    \label{fig:summary_task}

\end{figure}

To ensure that the Summary Task Component can generate the required summary-type sections accurately after prompting the LLM, we designed a prompt template specifically for the Summary Task. As shown in Figure \ref{fig:summary_task}, this prompt template consists of two main parts.

The first part involves Role Specification (Figure \ref{fig:summary_task}\circled{a}), which instructs the LLM to assume the role of a requirement assistant responsible for generating content for various summary-type sections based on the provided natural language text. 

The second part of the prompt template explicitly lists the sections for which content needs to be generated, along with detailed descriptions or definitions of the expected content for each section (Figure \ref{fig:summary_task}\circled{b}). For example, we provide the description for the Stakeholders/Users Section: ``Write who the product is intended to serve". Additionally, to mitigate hallucinations in the output of LLM, we specify that each stakeholder or user mentioned must be accompanied by an annotation indicating the text source from the provided natural language input. Similarly, the Glossary of Terms Section defines its content as: ``The glossary provides specific definitions of important terms used throughout the software requirements document".

Furthermore, for the second part of the prompt template, in addition to providing explicit definitions for each section, researchers have found that including guiding questions can further help guide the LLM to produce more relevant and structured content—for example, as illustrated in Figure \ref{fig:summary_task}\circled{b} for the Introduction section. If users find that providing only definitions does not yield satisfactory outputs, they may instead design guiding questions based on the expected content, enabling the LLM to generate results that better align with their intentions.

The Summary Task Component also includes a command list containing entries such as ``Write Introduction Section" and ``Write Stakeholders/Users Section." The Summary Task Component iterates through this command list, sequentially extracting commands and combining them with the prompt template before prompting the LLM. This process allows the LLM to focus on generating one section at a time, and the component populates each section into the SRS template accordingly.

Both the prompt template and the command list are designed to be extensible. If a required summary-type section is not initially included in the prompt template, users can extend the second part of the prompt template by adding relevant instructions in the same format. Likewise, by adding a corresponding command to the command list. This flexibility ensures that \reqinone remains adaptable to different SRS template structures and evolving requirements engineering needs.

\subsection{Requirement Extraction Task Component}
\begin{figure}[htbp]
    \centering
    \includegraphics[width=0.6\linewidth]{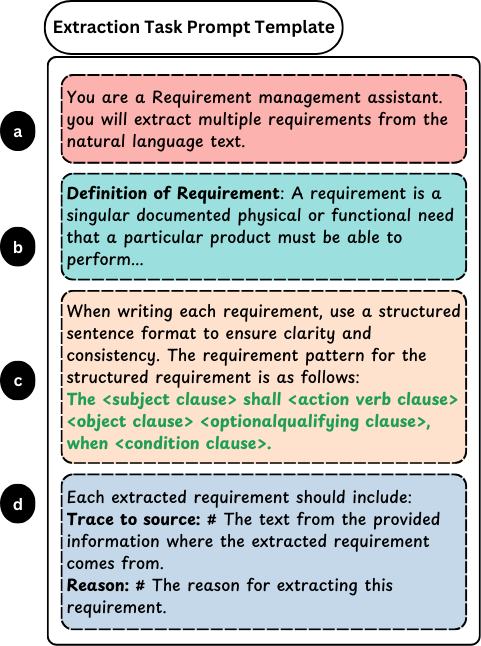}
    \caption{Prompt Template for Requirement Extraction Task}
    \label{fig:extraction_task}

\end{figure}
Similar to the Summary Task Component, we designed a specialized prompt template for the Requirement Extraction Task Component to facilitate prompting the LLM. The structure of this prompt template is illustrated in Figure \ref{fig:extraction_task} and consists of four main parts.

The first part is still Role Specification (Figure \ref{fig:extraction_task}\circled{a}), where the LLM is instructed to analyze the provided natural language text and extract relevant requirements. 

The second part is Requirement Definition (Figure \ref{fig:extraction_task}\circled{b}), which provides the LLM with a clear definition of what a requirement is. By explicitly defining requirements within the prompt, the LLM gains a precise understanding of the target content it needs to extract from the natural language text, thereby improving extraction accuracy.

The third part is Requirements Pattern (Figure \ref{fig:extraction_task}\circled{c}). In this part, the LLM is directed to extract requirements that conform to the following standardized format: “The $<$subject clause$>$ shall $<$action verb clause$>$ $<$object clause$>$ $<$optional qualifying clause$>$, when $<$condition clause$>$.” This format is derived from INCOSE documentation \cite{INCOSE} and serves as a guideline for the LLM, ensuring that all extracted requirements are expressed in a consistent and well-structured format. The purpose of enforcing this structured format is to promote clarity, consistency, and testability in requirement expressions. Additionally, structured format phrasing helps reduce ambiguity, making the requirements easier to understand, validate, and trace throughout the software development lifecycle \cite{incose2023incose}.

The fourth part includes Trace to Source (Figure \ref{fig:extraction_task}\circled{d}), where we instruct the LLM to append the source of each extracted requirement along with the reason for its extraction. This step is crucial in preventing hallucinations by compelling the LLM to justify the rationality of each requirement extraction based on trace to source before generating it. By linking each requirement to its original text source, we establish traceability within the SRS, ensuring that every requirement can be verified and traced back to its origin.

Once the Requirement Extraction Task Component provides the natural language text and the structured prompt template to the LLM, the LLM processes the input, analyzes the content, and returns a structured list of extracted requirements. The Requirement Extraction Task Component then organizes these extracted requirements into a requirements list, which is passed to the Requirement Classification Task Component for further classification and refinement.

\subsection{Requirement Classification Task Component}
In line with the design of the other components, we designed a specialized prompt template for the Requirement Classification Task Component. Upon receiving the requirement list from the Requirement Extraction Task Component, the Requirement Classification Task Component utilizes this prompt template to prompt the LLM. This prompt template transforms the LLM into a reasoning model specialized in classifying requirements into functional and non-functional categories.

\begin{figure}[htbp]
    \centering
    \includegraphics[width=0.6\linewidth]{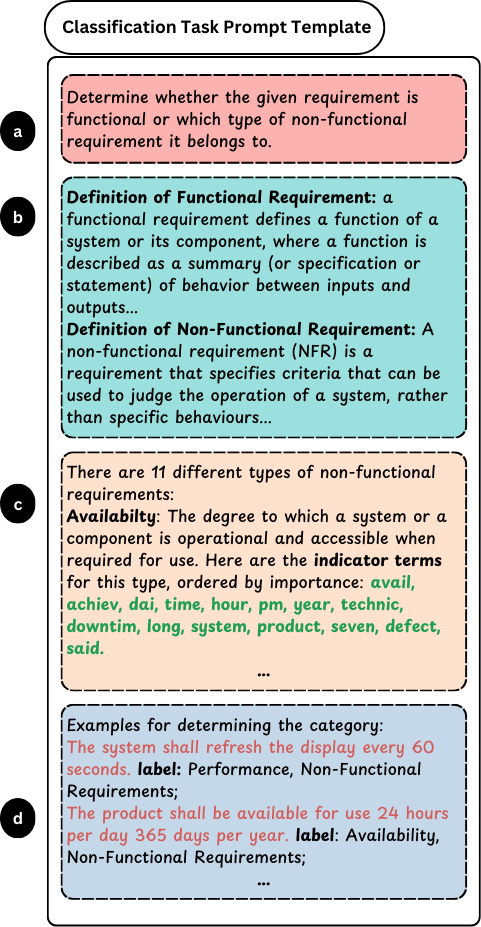}
    \caption{Prompt Template for Requirement Classification Task}
    \label{fig:classification_task}

\end{figure}

As illustrated in Figure 4, the prompt template consists of four main parts. The first part is Task Specification (Figure \ref{fig:classification_task}\circled{a}), which explicitly instructs the LLM to classify each requirement into either functional or non-functional categories. The second part is Definition of FRs and NFRs (Figure \ref{fig:classification_task}\circled{b}), providing a clear classification standard by defining FRs and NFRs to help the LLM distinguish between these categories.

The third part, Detailed Classification of NFRs, explains various subtypes within NFRs (Figure \ref{fig:classification_task}\circled{c}). This part defines 11 different subtypes of NFRs, such as Availability Requirement, Legal Requirement, and Maintainability Requirement, each accompanied by a corresponding definition. 
Additionally, Cleland-Huang et al.~\cite{cleland2007automated} proposed that different NFR types are often associated with specific keywords, which we refer to as indicator terms. These indicator terms are incorporated into the prompt template  following the corresponding requirement definitions, helping guide the LLM toward more accurate classification.

In the fourth part (Figure \ref{fig:classification_task}\circled{d}), we adopt the few-shot learning approach \cite{brown2020language}, where a set of labeled requirement examples is provided to enhance classification accuracy. These examples cover all 11 NFR subtypes and also include representative samples of FRs, ensuring that the LLM learns from diverse cases to improve its classification ability.

Considering that there is still no consensus in the software engineering community on the concept of NFRs \cite{glinz2007non,eckhardt2016non}, the categorization of NFRs may vary across different projects. NFRs in some cases may extend beyond the 11 NFR subtypes in the third part of the prompt template. To address this challenge, the prompt template is designed to be extensible. Users can customize the template by adding, modifying, or removing requirement category definitions based on their specific project needs or the adopted SRS template. Additionally, users can expand the few-shot learning examples by introducing new labeled requirements that align with the format used in the prompt template.

Once the LLM classifies all requirements in the list, an appropriate category label will be appended to each requirement. If a requirement is classified as NF, the label specifies which subtype of NFR it belongs to. The labeled requirements are then returned to the Requirement Classification Task Component, which organizes and populates them into the FRs Section or NFRs Section of the SRS template accordingly.

\subsection{Summary and Insights}
Unlike previous studies that typically instruct LLMs to generate the entire SRS in a single step, we introduces a novel strategy that decomposes the SRS generation task into three comparatively simpler sub-tasks: the Summary Task, the Requirement Extraction Task, and the Requirement Classification Task. Based on this strategy, we developed \reqinone, which composed of three corresponding components—each responsible for one sub-tasks. Every component is guided by a designed prompt template, tailored specifically for its corresponding task, and we adopt a zero-shot prompting approach to design these prompt templates \cite{radford2019language}.

A key advantage of our approach lies in its high degree of customizability. Users can freely adapt \reqinone to their preferred SRS format by modifying the contents of each prompt template. For instance, if a chosen SRS template includes a summary-type section not originally present in the Summary Task prompt template, the user can easily add the necessary description in the appropriate part of the prompt template. Conversely, if certain sections included in the original prompt template are irrelevant to the chosen SRS template, they can simply be removed.

Moreover, the use of prompt templates brings sustainability and extensibility to \reqinone. Users can continuously improve the output quality by adjusting the content of the prompt templates—for example, by modifying the definition of requirements in the requirement extraction prompt template, or by inserting more representative examples into the example part of the requirement classification prompt template. These kinds of adjustments allow \reqinone to be iteratively optimized over time, making its performance increasingly aligned with user expectations and domain-specific needs.

\section{Evaluation}
Given that our approach decomposes the process of generating an SRS into multiple subtasks, we focus on several key aspects when evaluating \reqinone: the overall quality of the generated SRS, the quality of the requirements within the SRS, and whether the requirements are correctly categorized into their appropriate sections. To this end, we propose the following three research questions:

\textbf{RQ1:} How does the overall quality of SRSs generated by \reqinone using different LLMs compare to SRSs produced by existing automated SRS generation methods and those written by entry-level requirements engineers?

\textbf{RQ2:} How does the quality of requirements generated by \reqinone compare to those from existing automated methods and entry-level engineers?

\textbf{RQ3:} How well does \reqinone perform in the requirement classification?

\subsection{Evaluation Design: User Study and Classification Task}
RQ1 and RQ2 were addressed through a survey-based evaluation using a questionnaire, while RQ3 was addressed via a classification task using benchmark datasets.

The questionnaire consisted of two parts. \textbf{Part 1, addressing RQ1,} included five evaluation parameters drawn from prior literature \cite{krishna2024using,wiegers1999writing,davis1993identifying}: Internal Consistency, Non-redundancy, Completeness, Conciseness, and Traceability. Participants read both the source text and the corresponding SRS before rating each parameter on a 1–5 Likert scale, where 1 indicates strong disagreement and 5 indicates strong agreement that the SRS meets the parameter. The aggregated scores were used to assess the overall quality of each SRS.

\textbf{Part 2, for RQ2,} followed a similar structure. Five requirements were randomly sampled from each SRS (ensuring one per category when possible). Each requirement was rated on five parameters: Unambiguous, Understandable, Correctness, Verifiable, and Conforming—also using a 1–5 Likert scale. As in RQ1, aggregated scores provided a quality assessment for each requirement.

Three software engineering experts participated in the study. Each evaluated five different SRSs: (1) SRS generated by \reqinone using ChatGPT-4o; (2) SRS generated by \reqinone using Llama3 (Version: Meta Llama3.1-8B); (3) SRS generated by \reqinone using DeepSeek-R1 (Version: DeepSeek-R1-0528-Qwen3-8B); (4) SRS generated by baseline, which directly uses GPT-4 to generate the SRS\cite{krishna2024using}; (5) SRS written by an entry-level requirements engineer \cite{krishna2024using}. To ensure unbiased evaluations, all participants were unaware of how each SRS was generated and had no prior involvement in this research.

\textbf{To address RQ3,} we evaluated \reqinone's classification component using the PROMISE dataset \cite{PRMOISE}, where the task involved classifying requirements as functional or non-functional and further categorizing NFRs into specific subtypes. To test generalizability, we constructed a new dataset—ReqFromSRS—by manually extracting requirements from the PURE dataset \cite{ferrari2017pure}. We compared performance against the NoRBERT baseline \cite{hey2020norbert}, using precision, recall, and F1 score as evaluation metrics.

\paragraph*{ReqFromSRS Dataset}
The PURE dataset is a collection of 79 SRS documents gathered from the web \cite{ferrari2017pure}. To evaluate the performance of \reqinone's requirement classification and its generalizability, we manually extracted 100 FRs and 100 NFRs from the SRSs in the PURE dataset.
\begin{itemize}
    \item Among the 100 NFRs, there were 10 Usability Requirements (US), 21 Performance Requirements (PE), 24 Security Requirements (SE), 12 Availability Requirements (A), 12 Maintainability Requirements (MN), 7 Portability Requirements (PO), 4 Scalability Requirements (SC), 8 Look \& Feel Requirements (LF), and 2 Legal Requirements (L). 
    \item During the manual extraction process, we only selected FRs explicitly stated under the FRs section of the SRS and labeled them with \textbf{F}. Similarly, for NFRs, we only extracted those explicitly assigned a NFR subtype within the SRS and labeled them accordingly. 
\end{itemize}


\begin{figure}[h]
    \centering
    \includegraphics[width=0.8\linewidth]{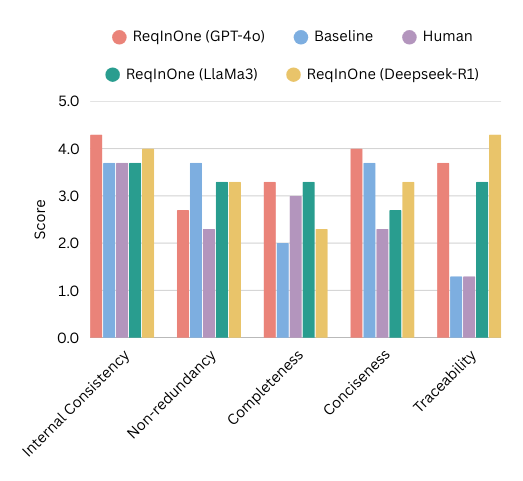}
    \caption{Overall evaluation of the five SRSs across five quality parameters. Each score represents the average rating from experts.}
    \label{fig:overall_SRS}
\end{figure}

\subsection{{RQ1: Overall Quality of SRSs}} 

As illustrated in Figure \ref{fig:overall_SRS}, the SRS generated by \reqinone using GPT-4o, despite receiving a relatively low score in Non-redundancy, achieved the highest scores in Internal Consistency, Completeness, Conciseness, and Traceability. This indicates that \reqinone (GPT-4o) delivers the best overall SRS quality among all evaluated methods.

Compared with the human-written SRS, \reqinone (GPT-4o) consistently outperformed across all five evaluation parameters. In terms of Internal Consistency and Completeness, the human-written SRS scored 3.8 and 3.0, whereas \reqinone (GPT-4o) achieved 4.2 and 3.2 respectively indicating that \reqinone (GPT-4o) can now generate more logically coherent and coverage-complete specifications than entry-level requirements engineer in many cases.

When compared to the baseline, which used GPT-4, \reqinone (GPT-4o) also showed superior performance in four out of five parameters, especially Traceability, where \reqinone achieved a significantly higher score. This is particularly notable given that GPT-4—the model behind the baseline—is approximately 12 times more expensive to use than GPT-4o. This comparison not only confirms the strong performance of \reqinone, but also highlights its cost-effectiveness, offering better results at a low computational cost. This suggests that our proposed strategy of decomposing the SRS generation into subtasks can effectively enhance the performance of LLMs in SRS generation.

Regarding traceability, both the baseline and human-written SRS did not clearly show traceability, receiving the lowest scores in this parameter. In contrast, all three SRSs generated by \reqinone clearly maintained traceability, with Deepseek-R1 performing best. 

Although the \reqinone powered by LLaMA3 and Deepseek-R1 did not achieve the highest overall scores, they did exhibit strengths in specific areas. Both outperformed the GPT-4o regarding Non-redundancy, suggesting that they may demonstrate stronger capability in extracting requirements.

\begin{tcolorbox}
\textbf{Answer to RQ1:} 
\reqinone (GPT-4o), delivers the highest overall SRS quality among all evaluated parameters, outperforming both the human-written SRS and the baseline, while maintaining low computational cost. Additionally, LLLaMA3 and Deepseek-R1 also showed strengths in Non-redundancy, suggesting potential in requirement extraction.
\end{tcolorbox}

\subsection{{RQ2: Quality of Generated Requirements}} 
\begin{figure}[h]
    \centering
    \includegraphics[width=0.8\linewidth]{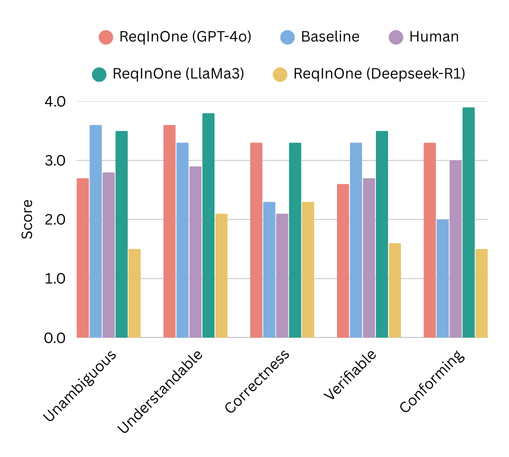}
    \caption{Evaluation of requirement quality within the generated SRS documents. Each parameter score represents the average rating by experts on five selected requirements from each SRS.}
    \label{fig:requirement_quality}
\end{figure}

Figure \ref{fig:requirement_quality} presents the evaluation of requirement quality in the generated SRSs. Overall, \reqinone using LLaMA3 produced the highest-quality requirements, while ReqInOne with Deepseek-R1 performed the worst.

Although \reqinone with LLaMA3 scored slightly lower than the baseline in the Unambiguous parameter, the difference was minimal. Both \reqinone (GPT-4o) and \reqinone (LLaMA3) achieved high scores in the Conforming parameter, outperforming the baseline. This improvement comes from using a requirement pattern in the requirement extraction prompt template, which guided LLMs to follow a consistent structure when generating requirements.

Although LLaMA3 performed better than GPT-4o in avoiding ambiguous phrasing and generating more easily understandable requirements, ambiguity remains a common challenge across all LLMs. These models often introduce unnecessary modifiers or redundant sentences, which can lead to vague or overly verbose requirements. Among the models evaluated, GPT-4o appeared to struggle with this issue the most. For instance, some requirements generated by GPT-4o included subjective terms such as ``appropriate" and ``user-friendly", which are highly subjective and can introduce ambiguity. This observation is further supported by the evaluation results shown in Figure \ref{fig:overall_SRS} (parameter: Non-redundancy) and Figure \ref{fig:requirement_quality} (parameter: Unambiguous), both of which reflect lower scores for GPT-4o in these aspects. Therefore, future research could explore fine-tuning LLMs to reduce the use of highly subjective terms, which may lead to improved requirement quality.

Regarding Correctness, the baseline fell clearly behind \reqinone (LLaMA3) and \reqinone (GPT-4o), generating more requirements that were not grounded in the source text. This highlights the effectiveness of the Trace to Source part in the requirement extraction prompt template, which helps reduce hallucinations and ensures better alignment with the source content.
\begin{tcolorbox}
\textbf{Answer to RQ2:} \reqinone, particularly with LLaMA3, generated higher-quality and more consistent requirements than both the baseline and human-written SRSs, benefiting from well-designed structured prompt templates that improved clarity, correctness, and conformity.
\end{tcolorbox}
\begin{table}[h] 
    \centering
    \caption{F/NFR classification results on PROMISE dataset across different tools}
    \resizebox{\columnwidth}{!}{ 
    \begin{tabular}{lccc|ccc}
        \toprule
        & \multicolumn{3}{c}{FR} & \multicolumn{3}{c}{NFR} \\
        \cmidrule(lr){2-4} \cmidrule(lr){5-7}
        Tool & P & R & F1 & P & R & F1 \\
        \midrule
        NoRBERT (Baseline) & \textbf{.92} & .88 & \textbf{.90} & .92 & \textbf{.95} & \textbf{.93} \\
        \reqinone (GPT-4o)  & .87 & \textbf{.95} & \textbf{.90} & \textbf{.96} & .90 & \textbf{.93} \\
        \reqinone (LlaMa3)  & .75 & .83 & .78 & .87 & .81 & .84 \\
        \reqinone (Deepseek-R1)  & .76 & .86 & .80 & .89 & .81 & .85 \\
        \bottomrule
    \end{tabular}
    }
    \label{tab:F/NFR_PROMISE}
\end{table}

\begin{table*}[h]
\centering
\caption{Classification of NFR subtypes on PROMISE dataset.}
\resizebox{\textwidth}{!}{
\begin{tabular}{lcccc}
\toprule
\textbf{Tool} 
& \textbf{NoRBERT} 
& \textbf{\reqinone (GPT-4o)} 
& \textbf{\reqinone (LLaMa3)} 
& \textbf{\reqinone (Deepseek-R1)} \\
\midrule
\rowcolor{gray!10}
& \textbf{\scriptsize P / R / F1} & \textbf{\scriptsize P / R / F1} & \textbf{\scriptsize P / R / F1} & \textbf{\scriptsize P / R / F1} \\
\midrule
A   & .80 / .76 / .78 & \textbf{.84} / \textbf{1} / \textbf{.91} & .30 / .90 / .45 & .77 / .95 / .85 \\
FT  & .60 / .60 / .60 & \textbf{.67} / \textbf{.80} / .73 & .62 / .50 / .56 & .78 / .70 / \textbf{.74} \\
L   & \textbf{.91} / .77 / \textbf{.83} & .55 / \textbf{.85} / .67 & .52 / .85 / .65 & .60 / .69 /.64 \\
LF  & .81 / .79 / .80 & \textbf{.91} / \textbf{.84} / \textbf{.88} & .78 / .76 / .77 & .83 / .63 / .72 \\
MN  & .62 / .47 / .53 & \textbf{.69} / \textbf{.65} / \textbf{.67} & .69 / .65 / .67 & .58 / .65 / .61 \\
O   & .78 / \textbf{.84} / \textbf{.81} & \textbf{.79} / .53 / .63 & .79 / .44 / .56 & .54 / .48 / .51 \\
PE  & \textbf{.92} / \textbf{.87} / \textbf{.90} & .87 / .83 / .85 & .79 / .83 / .81 & .81 / .81 / .81 \\
SC  & .76 / \textbf{.76} / \textbf{.76} & .71 / .71 / .71 & .73 / .52 / .61 & \textbf{.79} / .52 / .63 \\
SE  & .90 / \textbf{.92} / .91 & \textbf{.98} / .88 / .93 & .98 / .74 / .84 & .98 / .89 / \textbf{.94} \\
US  & .83 / \textbf{.88} / .86 & \textbf{.92} / .82 / \textbf{.87} & .90 / .66 / .76 & .79 / .79 / .79 \\
\midrule
\rowcolor{gray!10}
\textbf{Weighted F1} 
& \textbf{0.82} 
& \textbf{0.81} 
& \textbf{0.71} 
& \textbf{0.74} \\
\bottomrule
\end{tabular}
}
\label{tab:subtypes_PROMISE}
\end{table*}

\subsection{RQ3: Requirement Classification Performance}
\paragraph{Performance on PROMISE Dataset for FR/NFR Classification} To evaluate the performance of \reqinone’s Requirement Classification Component, we first assess its ability to classify requirements as either functional or non-functional on the PROMISE dataset. As illustrated in Table \ref{tab:F/NFR_PROMISE}, \reqinone using GPT-4o achieves competitive results when compared to the NoRBERT baseline. For FR, GPT-4o achieves an F1 score of 0.90, equal to NoRBERT, with a slightly lower precision (0.87 vs. 0.92) but notably higher recall (0.95 vs. 0.88). For NFR, GPT-4o outperforms the baseline in precision (0.96 vs. 0.92) while achieving the same F1 score of 0.93.

Meanwhile, \reqinone powered by Llama3 and DeepSeek-R1 also performs reasonably well. Llama3 achieves F1 scores of 0.78 (FR) and 0.84 (NFR), while DeepSeek-R1 yields 0.80 (FR) and 0.85 (NFR). Though they do not reach the level of GPT-4o or NoRBERT, their results suggest the potential of using local LLMs for future research in requirement classification tasks.

\paragraph{Performance on PROMISE Dataset for NFR Subtype Classification} 

To further evaluate the classification capabilities of \reqinone, we focus on the Classification of NFR Subtypes using the PROMISE dataset. This dataset includes 11 NFR subtypes: Availability (A), Fault Tolerance (FT), Legal (L), Look \& Feel (LF), Maintainability (MN), Operational (O), Performance (PE), Portability (PO), Scalability (SC), Security (SE), and Usability (US). As shown in Table \ref{tab:subtypes_PROMISE}, \reqinone using GPT-4o achieves a weighted F1 score of 0.81, which is nearly on par with the NoRBERT baseline score of 0.82. More importantly, GPT-4o surpasses NoRBERT in several individual subtypes, including Availability, Fault Tolerance, Look \& Feel, and Maintainability in terms of precision, recall, and F1 score. This demonstrates that \reqinone understands certain non-functional subtypes more than NoRBERT.

The results of LLaMa3 and DeepSeek-R1 also indicate solid performance. Although their weighted F1 scores are lower than GPT-4o and NoRBERT, their results align with the earlier FR and NFR classification task and suggest local models remain viable options for requirement classification.

\begin{table}[h] 
    \centering
    \caption{F/NFR classification results on ReqFromSRS dataset across different tools}
    \resizebox{\columnwidth}{!}{ 
    \begin{tabular}{lccc|ccc}
        \toprule
        & \multicolumn{3}{c}{FR} & \multicolumn{3}{c}{NFR} \\
        \cmidrule(lr){2-4} \cmidrule(lr){5-7}
        Tool & P & R & F1 & P & R & F1 \\
        \midrule
        NoRBERT (Baseline) & .84 & .45 & .59 & .63 & \textbf{.92} & .74 \\
        \reqinone (GPT-4o)  & \textbf{.85} & \textbf{.87} & \textbf{.86} & \textbf{.87} & .85 & \textbf{.86} \\
        \reqinone (LlaMa3)  & .80 & .79 & .79 & .79 & .80 & .80 \\
        \reqinone (Deepseek-R1)  & .82 & .71 & .76 & .74 & .84 & .79 \\
        \bottomrule
    \end{tabular}
    }
    \label{tab:F/NFR_reqfromsrs}
\end{table}
\paragraph{Performance on ReqFromSRS Dataset for FR/NFR Classification} Since our baseline NoRBERT is trained specifically on the PROMISE dataset, we constructed a new dataset—ReqFromSRS—to provide a more fair comparison and to evaluate the generalizability of \reqinone in classifying requirements. We performed the same FR/NFR classification task on this new dataset. 

As shown in Table \ref{tab:F/NFR_reqfromsrs}, \reqinone using GPT-4o significantly outperforms NoRBERT across all evaluation metrics. It achieves the highest precision, recall, and F1 score for both FRs and NFRs, indicating its strong generalization capability to previously unseen data. Even when powered by local models such as LLaMa3 and DeepSeek-R1, \reqinone still outperforms NoRBERT. Both local models maintain a balanced performance with F1 scores of 0.76–0.80, surpassing NoRBERT, especially in FR recall, where NoRBERT performs poorly (0.45). Although NoRBERT achieves a relatively high recall (0.92) for NFRs, its NFR precision (0.63) and FR recall (0.45) are substantially lower, indicating that NoRBERT tends to classify most requirements as NFR.

\begin{tcolorbox}
\textbf{Answer to RQ3:} 
\reqinone demonstrates strong performance comparable to the NoRBERT baseline on the PROMISE dataset and exhibits significantly better generalization on the ReqFromSRS dataset. Even when using local models like LLaMA3 and DeepSeek-R1, \reqinone still offers promising classification performance.
\end{tcolorbox}


\section{THREATS TO VALIDITY}
\paragraph{Internal Validity} To minimize randomness in LLM outputs and obtain stable results, we set the temperature of all LLMs to 0. We also specify the exact versions of the LLMs used in the evaluation. These settings reduce the diversity of possible outputs and also enhance the reproducibility of our study. Additionally, when using local models such as LLaMa3 and DeepSeek-R1, we opted for their 8B parameter versions instead of larger alternatives. This decision was made to balance computational feasibility and evaluation time, but it may have limited the performance of these models.

\paragraph{Construct Validity} To answer RQ1 and RQ2, all three participants involved in the survey are experts in the field of software engineering. The meaning of each evaluation parameter in the questionnaire was clearly explained to ensure consistency. Nonetheless, human judgment is inherently subjective, and differences in individual understanding may have introduced scoring bias. We attempted to mitigate this by selecting the most representative parameters reported in existing literature to assess SRS and requirement quality.

\paragraph{Conclusion Validity} Instead of evaluating every requirement in the SRS, which would have imposed a heavy workload on the participants and possibly affected their judgment, we selected a sample of requirements that were as diverse as possible across different requirement types. This sampling strategy helps ensure coverage while maintaining evaluation quality, but it may still limit the comprehensiveness of our assessment.

\section{Conclusion}
In this paper, we proposed \reqinone, an LLM-based agent designed to automatically SRS by scheduling three tasks: summarization, requirement extraction, and requirement classification. Our evaluation shows that the SRS and individual requirements generated by \reqinone are of higher quality and more compliant with standard SRS guidelines than those produced by baseline methods or entry-level requirements engineers. Additionally, \reqinone achieves high accuracy and strong generalizability in the requirement classification task. These results demonstrate the potential of \reqinone to improve the efficiency of requirements engineering. 

As part of future work, we aim to extend \reqinone by incorporating automated requirement validation mechanisms, enabling a more robust generation–validation–update workflow to further improve the quality and reliability of generated SRSs.

\balance
\bibliographystyle{ieeetr}
\bibliography{all}

\begin{thebibliography}{10}

\bibitem{hofmann2001requirements}
H.~F. Hofmann and F.~Lehner, ``Requirements engineering as a success factor in
  software projects,'' {\em IEEE software}, vol.~18, no.~4, p.~58, 2001.

\bibitem{doe2011recommended}
J.~Doe, ``Recommended practice for software requirements specifications
  (ieee),'' {\em IEEE, New York}, 2011.

\bibitem{belfo2012people}
F.~Belfo, ``People, organizational and technological dimensions of software
  requirements specification,'' {\em Procedia Technology}, vol.~5,
  pp.~310--318, 2012.

\bibitem{VisualParadigm}
``\textit{VisualParadigm}.'' Available: \url{https://www.visual-paradigm.com/},
  2001.

\bibitem{reqview}
``\textit{ReqView}.'' Available: \url{https://www.reqview.com/}, 2015.

\bibitem{elemtool}
``\textit{Elementool}.'' Available: \url{https://www.elementool.com/}, 2000.

\bibitem{georgiades2010automatic}
M.~G. Georgiades and A.~S. Andreou, ``Automatic generation of a software
  requirements specification (srs) document,'' in {\em 2010 10th International
  Conference on Intelligent Systems Design and Applications}, pp.~1095--1100,
  IEEE, 2010.

\bibitem{chen2021evaluating}
M.~Chen, J.~Tworek, H.~Jun, Q.~Yuan, H.~P. d.~O. Pinto, J.~Kaplan, H.~Edwards,
  Y.~Burda, N.~Joseph, G.~Brockman, {\em et~al.}, ``Evaluating large language
  models trained on code,'' {\em arXiv preprint arXiv:2107.03374}, 2021.

\bibitem{kasneci2023chatgpt}
E.~Kasneci, K.~Se{\ss}ler, S.~K{\"u}chemann, M.~Bannert, D.~Dementieva,
  F.~Fischer, U.~Gasser, G.~Groh, S.~G{\"u}nnemann, E.~H{\"u}llermeier, {\em
  et~al.}, ``Chatgpt for good? on opportunities and challenges of large
  language models for education,'' {\em Learning and individual differences},
  vol.~103, p.~102274, 2023.

\bibitem{zhao2023survey}
W.~X. Zhao, K.~Zhou, J.~Li, T.~Tang, X.~Wang, Y.~Hou, Y.~Min, B.~Zhang,
  J.~Zhang, Z.~Dong, {\em et~al.}, ``A survey of large language models,'' {\em
  arXiv preprint arXiv:2303.18223}, 2023.

\bibitem{roziere2023code}
B.~Roziere, J.~Gehring, F.~Gloeckle, S.~Sootla, I.~Gat, X.~E. Tan, Y.~Adi,
  J.~Liu, T.~Remez, J.~Rapin, {\em et~al.}, ``Code llama: Open foundation
  models for code,'' {\em arXiv preprint arXiv:2308.12950}, 2023.

\bibitem{achiam2023gpt}
J.~Achiam, S.~Adler, S.~Agarwal, L.~Ahmad, I.~Akkaya, F.~L. Aleman, D.~Almeida,
  J.~Altenschmidt, S.~Altman, S.~Anadkat, {\em et~al.}, ``Gpt-4 technical
  report,'' {\em arXiv preprint arXiv:2303.08774}, 2023.

\bibitem{guo2025deepseek}
D.~Guo, D.~Yang, H.~Zhang, J.~Song, R.~Zhang, R.~Xu, Q.~Zhu, S.~Ma, P.~Wang,
  X.~Bi, {\em et~al.}, ``Deepseek-r1: Incentivizing reasoning capability in
  llms via reinforcement learning,'' {\em arXiv preprint arXiv:2501.12948},
  2025.

\bibitem{radford2019language}
A.~Radford, J.~Wu, R.~Child, D.~Luan, D.~Amodei, I.~Sutskever, {\em et~al.},
  ``Language models are unsupervised multitask learners,'' {\em OpenAI blog},
  vol.~1, no.~8, p.~9, 2019.

\bibitem{brown2020language}
T.~Brown, B.~Mann, N.~Ryder, M.~Subbiah, J.~D. Kaplan, P.~Dhariwal,
  A.~Neelakantan, P.~Shyam, G.~Sastry, A.~Askell, {\em et~al.}, ``Language
  models are few-shot learners,'' {\em Advances in neural information
  processing systems}, vol.~33, pp.~1877--1901, 2020.

\bibitem{wei2022chain}
J.~Wei, X.~Wang, D.~Schuurmans, M.~Bosma, F.~Xia, E.~Chi, Q.~V. Le, D.~Zhou,
  {\em et~al.}, ``Chain-of-thought prompting elicits reasoning in large
  language models,'' {\em Advances in neural information processing systems},
  vol.~35, pp.~24824--24837, 2022.

\bibitem{hey2020norbert}
T.~Hey, J.~Keim, A.~Koziolek, and W.~F. Tichy, ``Norbert: Transfer learning for
  requirements classification,'' in {\em 2020 IEEE 28th international
  requirements engineering conference (RE)}, pp.~169--179, IEEE, 2020.

\bibitem{surana2019intelligent}
C.~S. R.~K. Surana, D.~B. Gupta, S.~P. Shankar, {\em et~al.}, ``Intelligent
  chatbot for requirements elicitation and classification,'' in {\em 2019 4th
  International Conference on Recent Trends on Electronics, Information,
  Communication \& Technology (RTEICT)}, pp.~866--870, IEEE, 2019.

\bibitem{ronanki2022chatgpt}
K.~Ronanki, B.~Cabrero-Daniel, and C.~Berger, ``Chatgpt as a tool for user
  story quality evaluation: Trustworthy out of the box?,'' in {\em
  International Conference on Agile Software Development}, pp.~173--181,
  Springer, 2022.

\bibitem{luitel2024improving}
D.~Luitel, S.~Hassani, and M.~Sabetzadeh, ``Improving requirements
  completeness: Automated assistance through large language models,'' {\em
  Requirements Engineering}, vol.~29, no.~1, pp.~73--95, 2024.

\bibitem{ronanki2023investigating}
K.~Ronanki, C.~Berger, and J.~Horkoff, ``Investigating chatgpt’s potential to
  assist in requirements elicitation processes,'' in {\em 2023 49th Euromicro
  Conference on Software Engineering and Advanced Applications (SEAA)},
  pp.~354--361, IEEE, 2023.

\bibitem{ezzini2023ai}
S.~Ezzini, S.~Abualhaija, C.~Arora, and M.~Sabetzadeh, ``Ai-based question
  answering assistance for analyzing natural-language requirements,'' in {\em
  2023 IEEE/ACM 45th International Conference on Software Engineering (ICSE)},
  pp.~1277--1289, IEEE, 2023.

\bibitem{abdelfattah2023roadmap}
A.~M. Abdelfattah, N.~A. Ali, M.~Abd~Elaziz, and H.~H. Ammar, ``Roadmap for
  software engineering education using chatgpt,'' in {\em 2023 International
  Conference on Artificial Intelligence Science and Applications in Industry
  and Society (CAISAIS)}, pp.~1--6, IEEE, 2023.

\bibitem{el2023ai}
A.~El-Hajjami, N.~Fafin, and C.~Salinesi, ``Which ai technique is better to
  classify requirements? an experiment with svm, lstm, and chatgpt,'' {\em
  arXiv preprint arXiv:2311.11547}, 2023.

\bibitem{endres2024can}
M.~Endres, S.~Fakhoury, S.~Chakraborty, and S.~K. Lahiri, ``Can large language
  models transform natural language intent into formal method
  postconditions?,'' {\em Proceedings of the ACM on Software Engineering},
  vol.~1, no.~FSE, pp.~1889--1912, 2024.

\bibitem{leite2024extracting}
G.~Leite, F.~Arruda, P.~Antonino, A.~Sampaio, and A.~Roscoe, ``Extracting
  formal smart-contract specifications from natural language with {LLMs},'' in
  {\em International Conference on Formal Aspects of Component Software},
  pp.~109--126, Springer, 2024.

\bibitem{krishna2024using}
M.~Krishna, B.~Gaur, A.~Verma, and P.~Jalote, ``Using llms in software
  requirements specifications: an empirical evaluation,'' in {\em 2024 IEEE
  32nd International Requirements Engineering Conference (RE)}, pp.~475--483,
  IEEE, 2024.

\bibitem{chang2024survey}
Y.~Chang, X.~Wang, J.~Wang, Y.~Wu, L.~Yang, K.~Zhu, H.~Chen, X.~Yi, C.~Wang,
  Y.~Wang, {\em et~al.}, ``A survey on evaluation of large language models,''
  {\em ACM transactions on intelligent systems and technology}, vol.~15, no.~3,
  pp.~1--45, 2024.

\bibitem{fan2023large}
A.~Fan, B.~Gokkaya, M.~Harman, M.~Lyubarskiy, S.~Sengupta, S.~Yoo, and J.~M.
  Zhang, ``Large language models for software engineering: Survey and open
  problems,'' in {\em 2023 IEEE/ACM International Conference on Software
  Engineering: Future of Software Engineering (ICSE-FoSE)}, pp.~31--53, IEEE,
  2023.

\bibitem{ferrari2017pure}
A.~Ferrari, G.~O. Spagnolo, and S.~Gnesi, ``Pure: A dataset of public
  requirements documents,'' in {\em 2017 IEEE 25th international requirements
  engineering conference (RE)}, pp.~502--505, IEEE, 2017.

\bibitem{ReqInOne}
``\textit{ReqInOne}.'' Available: \url{https://github.com/TaohongZ/ReqInOne},
  2025.

\bibitem{tian2024evaluating}
X.~Tian, ``Evaluating the repair ability of llm under different prompt
  settings,'' in {\em 2024 IEEE International Conference on Software Services
  Engineering (SSE)}, pp.~313--322, IEEE, 2024.

\bibitem{ieee830-1998}
{{IEEE}}, ``{IEEE} recommended practice for software requirements
  specifications.'' IEEE Std 830-1998, 1998.
\newblock pp. 1--40.

\bibitem{INCOSE}
``\textit{Guide to Writing Requirements}.'' Available:
  \url{https://www.incose.org/docs/default-source/working-groups/requirements-wg/gtwr/incose\_rwg\_gtwr\_v4\_040423\_final\_dra\-
  fts.pdf\?sfvrsn=5c877fc7\_2}, 2023.

\bibitem{incose2023incose}
INCOSE, {\em INCOSE systems engineering handbook}.
\newblock John Wiley \& Sons, 2023.

\bibitem{cleland2007automated}
J.~Cleland-Huang, R.~Settimi, X.~Zou, and P.~Solc, ``Automated classification
  of non-functional requirements,'' {\em Requirements engineering}, vol.~12,
  pp.~103--120, 2007.

\bibitem{glinz2007non}
M.~Glinz, ``On non-functional requirements,'' in {\em 15th IEEE international
  requirements engineering conference (RE 2007)}, pp.~21--26, IEEE, 2007.

\bibitem{eckhardt2016non}
J.~Eckhardt, A.~Vogelsang, and D.~M. Fern{\'a}ndez, ``Are" non-functional"
  requirements really non-functional? an investigation of non-functional
  requirements in practice,'' in {\em Proceedings of the 38th international
  conference on software engineering}, pp.~832--842, 2016.

\bibitem{wiegers1999writing}
K.~E. Wiegers, ``Writing quality requirements,'' {\em Software Development},
  vol.~7, no.~5, pp.~44--48, 1999.

\bibitem{davis1993identifying}
A.~Davis, S.~Overmyer, K.~Jordan, J.~Caruso, F.~Dandashi, A.~Dinh, G.~Kincaid,
  G.~Ledeboer, P.~Reynolds, P.~Sitaram, {\em et~al.}, ``Identifying and
  measuring quality in a software requirements specification,'' in {\em [1993]
  Proceedings First International Software Metrics Symposium}, pp.~141--152,
  Ieee, 1993.

\bibitem{PRMOISE}
C.-H. Jane, M.~Sepideh, L.~Huang, and P.~Dan, ``\textit{PRMOISE NFR Dataset}.''
  Available: \url{https://zenodo.org/records/268542}, 2007.

\end{thebibliography}

\end{document}